\documentclass[useAMS,usenatbib,usegraphicx]{mn2e}
\usepackage{natbib}
\usepackage{color}
\usepackage{mathrsfs}
\usepackage{times,xspace,amssymb}
\usepackage{url}
\usepackage{amsmath}
\usepackage{hyperref}
\hypersetup{colorlinks=True, linkcolor=red, citecolor=blue, anchorcolor=blue}
\DeclareMathAlphabet{\mathbi}{OT1}{ptm}{bx}{it}
\SetMathAlphabet\mathbi{bold}{OT1}{ptm}{bx}{it}
\def\sss{\scriptscriptstyle}

\def\bcd{\rm H\alpha^{b}/H\beta^{b}}
\def\bnd{\rm H\alpha^{n}/H\beta^{n}}

\def\sss{\scriptscriptstyle}
\def\bhm{M_{\bullet}}

\def\ergs{\rm erg~s^{-1}}
\def\fblr{f_{\sss\rm BLR}}

\def\mathdotM{\dot{\mathscr{M}}}
\def\rblr{R_{\sss{\rm BLR}}}
\def\sunm{M_\odot}
\def\tauhb{\tau_{_{\rm H\beta}}}

\def\mathdotM{\dot{\mathscr{M}}}

\def\mnras{MNRAS}
\def\apj{ApJ}
\def\aj{AJ}

\def\aap{A\&A}

\def\aplett{ApJ Letters}
\def\apjs{ApJS}

\def\araa{ARA\&A}
\def\nat{Nature}

\def\pasp{Pub. Ast. Soc. Pacific}

\begin{document}

\voffset=-0.6in

\title[{Reddening of AGN}]{Reddening of the BLR and NLR in AGN From a Systematic Analysis of Balmer Decrement}

\author[Lu et al.]
{
Kai-Xing Lu$^{1,2}$\thanks{E-mail: lukx@ynao.ac.cn}, 
Yinghe Zhao$^{1,2}$\thanks{E-mail: zhaoyinghe@ynao.ac.cn}, 
Jin-Ming Bai$^{1,2}$\thanks{E-mail: baijinming@ynao.ac.cn}, 
Xu-Liang Fan$^{3}$\\
$^{1}$Yunnan Observatories (YNAO), Chinese Academy of Sciences, Kunming 650011, Yunnan, China \\ 
$^{2}$Key Laboratory for the Structure and Evolution of Celestial Objects, Chinese Academy of Sciences, Kunming 650216, China \\
$^{3}$School of Physics, Huazhong University of Science and Technology, Wuhan 430074, China
}

\pagerange{\pageref{firstpage}--\pageref{lastpage}} \pubyear{2018} 
\maketitle
\label{firstpage}

\begin{abstract}
We selected an active galactic nuclei (AGN) sample ($0 < z \le 0.35$) 
from Sloan Digital Sky Survey Data Release 7, 
and measured the broad- ($\bcd$) and narrow-line Balmer decrements ($\bnd$) of 554 selected AGNs. 
We found that the distributions of Balmer decrements can be fitted by a Gaussian function 
and give the best estimates of $\bcd = 3.16$ with a standard deviation 0.07 dex, 
and $\bnd = 4.37$ with a standard deviation 0.10 dex. 
We inspected the distributions of $\bcd$ and $\bnd$ in the Baldwin$-$Phillips$-$Terlevich (BPT) diagram 
and found that only narrow-line Balmer decrements depend on 
the physical conditions of the narrow-line region (NLR). 
We tested the relationship between $\bcd$ and $\bnd$, 
and found that $\bcd$ does not correlate with $\bnd$. 
We investigated the relationship between Balmer decrements and Seyfert sub-type, 
and found that only broad-line Balmer decrements correlate with Seyfert sub-type, 
We also examined the dependency of Balmer decrements on AGN properties, 
and found that Balmer decrements have no correlation with optical luminosity, but show some dependence on accretion rate.  
These results indicate that the NLR is subject to more reddening by dust than the broad-line region (BLR). 
\end{abstract} 


\section{Introduction}
The unified model of active galactic nuclei (AGN) proposes that the variety in AGN type 
is just the result of viewing from different orientations (\citealt{Antonucci1993,Urry1995}). 
A dusty toroidal structure (i.e., the torus) surrounding an accreting massive black hole (BH) 
is believed to be a common feature of AGN. 
The broad-line region (BLR) is close to the central region and is believed to be located within the inner radius of the dusty torus. 
The narrow-line region (NLR) is located farther from the black hole and accretion disc, at the distance of a few pc. 
Many hard X-ray surveys have showed that $\sim$70\% of all local AGNs are obscured, 
where the obscuration in X-ray regime is produced by multiple absorbers mostly associated with the torus and the BLR, 
the obscuring material in the infrared is a transition zone between the BLR and NLR (e.g., \citealt{Burlon2011,Ramos2017}).  
In the optical regime, \cite{Gaskell2017arXiv} suggests that outflowing dusty clumps driven by radiation pressure acting on the dust 
complicate the study of the geometric structure and kinematics of the BLR and the search for sub-parsec supermassive black hole (SMBH) binaries. 

The internal reddening caused by obscuring material attenuates the activity of AGN and blurs the nature of AGN, 
such as the underestimate of luminosity and accretion rate. For example, 
The reverberation mapped spectra of IRAS F12397+333 ($z=0.0435$) 
have a very red colour resulting from the internal reddening of AGN (Galactic extinction corrected; \citealt{Du2014}). 
After correcting internal reddening, 
the luminosity at 5100\AA  ~and dimensionless accretion rate increase 4 and 9 times, respectively (\citealt{Du2014,Hu2015}). 
In addition, the internal reddening in AGN, which always produced by dust in the local of AGN or the host galaxy, 
are useful for us to understand the accretion and feedback process of AGN. 

Using the Balmer decrement as an indicator of reddening, 
many works have studied the internal reddening of AGN in detail  for different AGN samples 
(e.g., \citealt{Costero1977,Gaskell1982,Gaskell1984,Gaskell2004,Dong2005,LaMura2007,
Dong2008,Heard2016,Baron2016,Gaskell2017}).  
For the BLR, \cite{Gaskell2017} suggested that the intrinsic Balmer 
decrement in extremely blue AGN is $\bcd\approx2.72\pm0.04$\footnote
{The superscript $^{\rm b}$ and $^{\rm n}$ in the text stand for the broad emission-line and narrow emission-line, respectively}, 
which is consistent with the Baker-Menzel Case B value of 2.74 \citep{Osterbrock2006}. 
For the NLR, 
the best overall average value of 3.1 is adopted for the intrinsic $\bnd$ ratio \citep{Gaskell1982,Gaskell1984,Wysota1988,Heard2016}, 
which is a bit larger than the recombination value of 2.85 resulting from the effects of collisional excitation \citep{Osterbrock2006}. 

However there still exist some open questions including 
(1) many works show that the distributions of the broad-line Balmer decrement cover a wide range, 
but the median or average value varies from one sample to another. 
For example, \cite{LaMura2007} derived $\bcd=3.45\pm 0.65$, 
\cite{Dong2008} gave $\bcd=3.06$ with standard deviation 0.03 dex, 
\cite{Gaskell2017} suggested that the intrinsic $\bcd$ is 2.72, 
\cite{Zhou2006} found the mean $\bcd$ ratio to be 3.028 with a dispersion of 0.36, etc. 
(2) \cite{deZotti1985} and \cite{Heard2016} found that broad-line Balmer decrements 
are greater than narrow-line Balmer decrements, which is opposite to the result of \cite{Baron2016}; 
(3) The nature and distribution of dust causing internal reddening of AGN remain unknown 
(e.g., \citealt{Heard2016, Baron2016,Gaskell2017}), 
for example whether dust causing reddening in the BLR and continuum also causes in the NLR or not.  
To investigate these issues, in this paper, we study the properties of broad- and narrow-line Blamer decrements 
based on a AGN sample selected from the Sloan Digital Sky Survey Data Release 7 (SDSS DR7; \citealt{York2000,Schneider2010}). 
In Section~\ref{sec-sample}, we describe the construction of sample. 
We fit spectra and obtain the measurements of emission-line fluxes  in Section~\ref{sec-fit}, 
and give results in Section~\ref{sec-r}. 
Section~\ref{sec-d} is discussion. We draw our conclusions in Section~\ref{sec-c}. 

\section{Sample Construction}
\label{sec-sample}
Based on the H$\beta$ and H$\alpha$ emission line, in this paper, 
we concentrate on studying the properties of Balmer decrements 
and exploring potential difference (if existences) between broad- and narrow-line Balmer decrements. 
However, it is hard to separate the NLR Balmer lines from the BLR Balmer lines, 
since narrow lines sometimes blend with broad lines \citep{Dong2008,Hu2008,Stern2012a,Stern2012b,Stern2013}. 
To this purpose, we selected a sample of broad-line AGNs from 
the Sloan Digital Sky Survey Data Release 7 \citep{Schneider2010} using the following criteria: 

\begin{enumerate}
 \item 
 We limited redshifts $\le 0.35$ at first so that the H$\alpha$ line has a sufficient number of good spectral pixels 
 within the SDSS spectral wavelength coverage, which gives a total sample consisting of 3532 AGNs. 

\item
Then we selected objects with spectral signal-to-noise ratio (S/N)$>$15 
to limit the complex model well in spectral fitting scheme (see Section~\ref{sec-fit}). 
Therefore, 1005 AGNs are excluded from parent sample.

\item 
In order to ensure reliable measurement of narrow emission-line, 
we further selected objects that the signal-to-noise ratio of the peak of narrow emission-lines 
(including H$\beta^{\rm n}~\lambda$4861,  H$\alpha^{\rm n}~\lambda$6563 and [N~II]~$\lambda$6584) 
greater than 10, and found only 616 AGNs satisfy with this criteria. 

 \item 
 We excluded those objects with apparent absorption lines which indicates significant contamination from host galaxies. 
 Finally, we selected 554 AGNs. 
\end{enumerate}

\begin{figure}
\centering
\includegraphics[angle=0,width=0.5\textwidth]{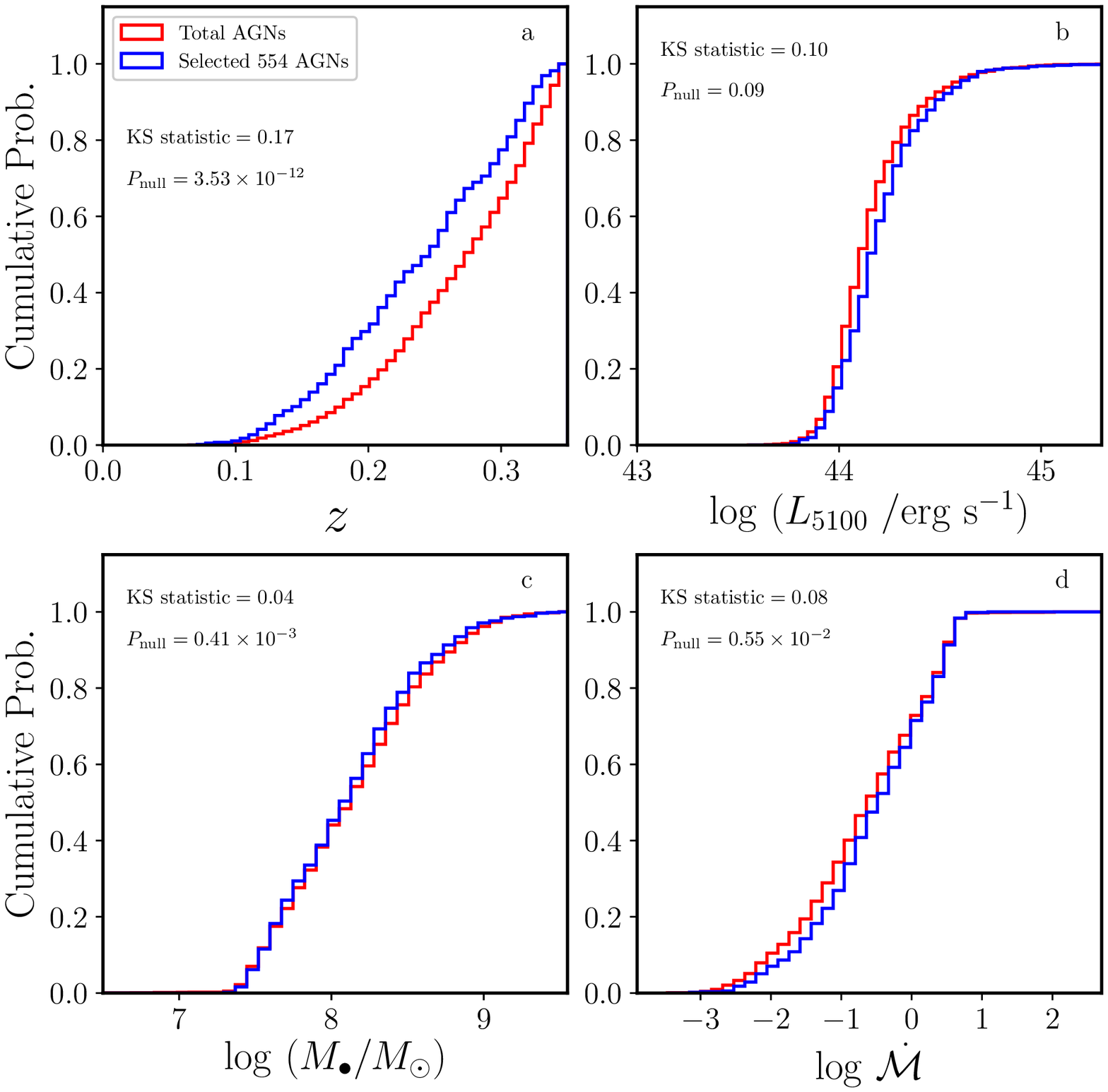}
\caption{
Two-sample Kolmogorov$-$Smirnov (KS) test. 
The cumulative distribution function in red  and blue correspond to the total sample and selected sample (see Section~\ref{sec-sample}). 
KS statistics and null hypothesis are quoted in each panel. 
}
\label{fig-kstest}
\end{figure}

To check whether the selected sample bias from the parent sample in AGN properties, 
we plotted the cumulative distribution function of the primary AGNs parameters 
(calculated from the best fitted results, see Section~\ref{sec-fit}) 
for selected 554 AGNs and total sample in Figure~\ref{fig-kstest}. 
Then we employed Kolmogorov$-$Smirnov (KS) test to quantify the differences, 
the results are quoted in Figure~\ref{fig-kstest}. 
The test shows that the distribution of selected 554 AGNs are almost same with parent sample (except for redshift).  

\begin{figure*}
\centering
\includegraphics[angle=0,width=0.95\textwidth]{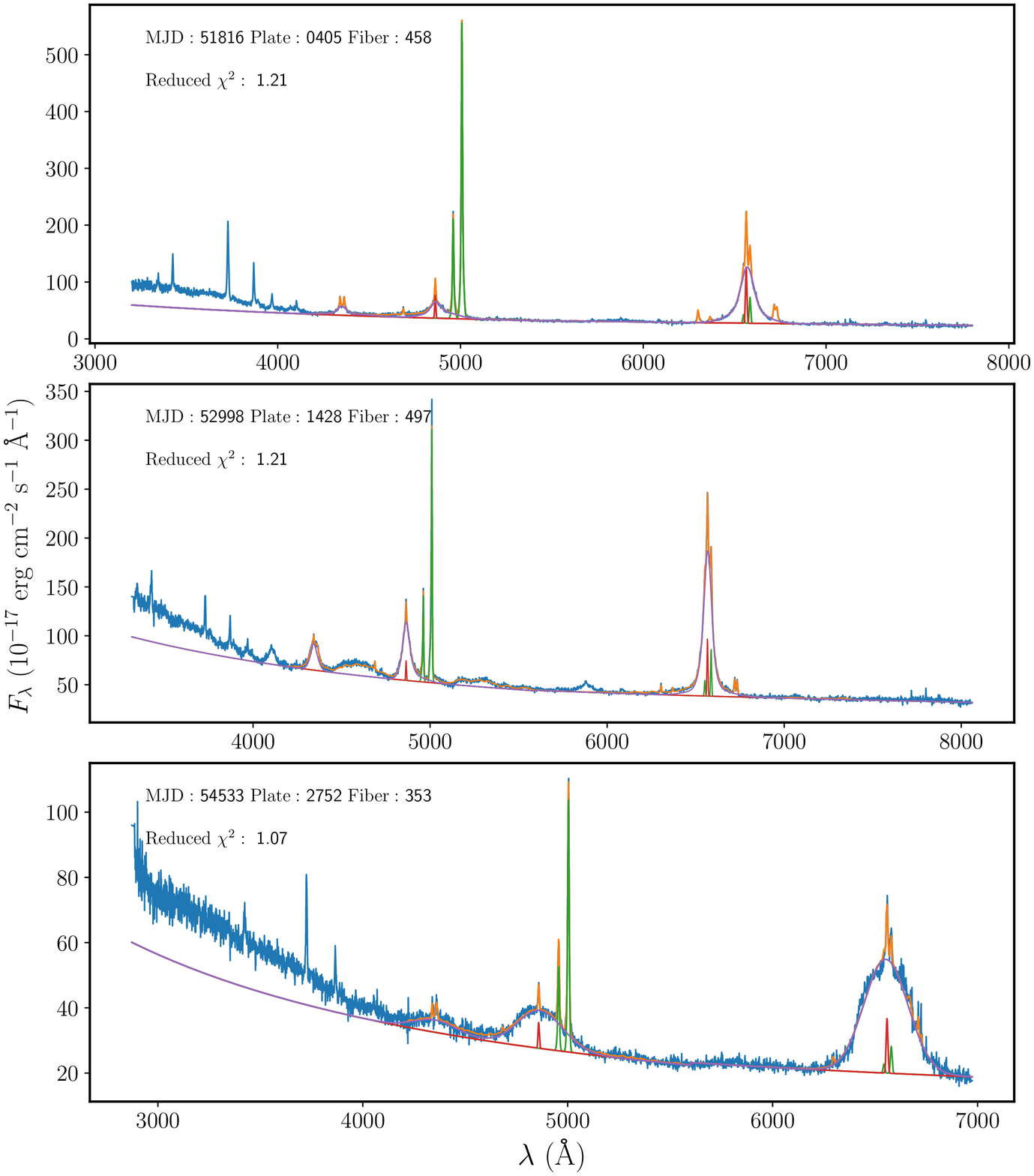} 
\caption{
Representative examples of multi-component fitting of the SDSS spectra. 
In each panel, we plot the the SDSS spectra corrected for Galactic extinction and redshift (blue). 
The sum of the best-fitting components is displayed in orange. 
The continuum, which is modelled as a broken power law with a break at 5600\AA, plus the broad Balmer lines, are shown in purple. 
Where the broad Balmer lines are fitted with 2, 3 and 4 Gaussian functions from top, middle to bottom panel, respectively.  
Narrow Balmer lines (H$\beta^{\rm n}~\lambda$4861 and  H$\alpha^{\rm n}~\lambda$6563) are plotted in red. 
The [O\,III] doublet $\lambda$5007/$\lambda$4959 and [N\,II] doublet 
$\lambda$6583/$\lambda$6548 are plotted in green. 
}
\label{fit-spec-fitting}
\end{figure*}

\section{Spectral Fitting and Measurements}
\label{sec-fit}
Spectral fitting scheme is widely used in the optical spectra study of AGNs (e.g., \citealt{Hu2008,Dong2008,Dong2011,Wang2009,Shen2011,Stern2013,Lu2016b}). 
Meanwhile, the disadvantage of spectral decomposition is also emphasized in many works \citep{Dong2008,Stern2013}. 
Briefly, the limitations and complexities of spectral decomposition attributes to the facts that 
there are essentially no emission-line free regions where the continuum can be determined well; 
Fe\,II emission lines and narrow corona lines are blended with broad lines; 
AGN continuum cannot be described by a single power law from red to blue side of spectrum, 
which means that we have to determine the local continuum for the H$\alpha$ and H$\beta$ regions separately (also see \citealt{Dong2008}). 
Therefore, we simultaneously fit the spectrum with multiple components, 
which include the continuum, the Fe II multiplets and several emission lines, 
giving emphasis on proper determination of the local pseudo-continua. 
This fitting scheme is similar to the process adopted by \cite{Dong2008}. 
In brief, main process of spectral fitting is described as follows: 

We corrected Galactic extinction and redshift for each SDSS spectrum, 
then we fitting spectrum in rest-frame wavelength range of 4200\AA~to 7500\AA~
using a broken power law with a break wavelength at 5600\AA. 
The optical Fe\,II emission lines are modelled by two separate sets of templates constructed by \cite{Dong2008, Dong2011} based on the measurements of I Zw 1 by \cite{Veron2004}, one for broad Fe\,II line system and the other for narrow Fe\,II line system. 
We modelled emission lines from H$\gamma$ to [S\,II] $\lambda6731
$\footnote{
Some emission-line regions in the fitting window are masked out, 
because they have no effect on the results.
They includes [Cl III] $\lambda5538$, He\,I $\lambda5876$ and [Ar III] $\lambda7136$. 
Some emission lines are not added to the fitting model, 
because either they are too weak to constrain in the fit or they have little effect on the results. 
They includes He\,I $\lambda4471$, [Fe\,VII] $\lambda5158$ and the $\lambda\lambda5721,6086$, 
[N\,I] $\lambda5200$, [Ca V] $\lambda5310$, He\,I $\lambda7066$ and [O\,II] $\lambda7320$.

} as follows. 
Broad hydrogen Balmer lines (H$\alpha^{\rm b}$, H$\beta^{\rm b}$, H$\gamma^{\rm b}$) are modelled with two to four Gaussians. 
The broad He\,II $\lambda4686$ line is modelled with one Gaussian.
[O\,III] $\lambda4363$ and the $\lambda\lambda4959,5007$ doublet
are assumed to have the same redshifts and profiles,
and each is modelled with one to two Gaussians. 
Other corona lines are modelled with one Gaussian. 
Narrow Balmer lines (H$\alpha^{\rm n}$, H$\beta^{\rm n}$, H$\gamma^{\rm n}$), [N\,II] and [S\,II] doublets
are assumed to have the same redshift and profile. 
The flux ratio of the [O\,III] doublet $\lambda$5007/$\lambda$4959 and [N\,II] doublet 
$\lambda$6583/$\lambda$6548 are fixed to the theoretical value of 2.96 (more details refer to \citealt{Dong2008}). 
In Figure~\ref{fit-spec-fitting}, we also over-plotted the fitted results, as shown by the orange line in each panel.  

From the best fitted results, we calculated the monochrome luminosity at 5100\AA~($L_{5100}$), 
the full width at half maximum (FWHM) of broad H$\beta$ emission line. 
Then the black hole mass ($M_{\bullet}$) is estimated  by 
\begin{equation}
\bhm=\fblr \frac{V_{\rm BLR}^2\rblr}{G}, 
\end{equation}
where $\rblr=c\tauhb$, $\tauhb$ estimated from $R_{\rm BLR}-L_{\rm 5100}$ relationship \citep{Bentz2013} 
is the H$\beta$ time lag with respective to the 5100~{\AA} continuum, 
$G$ is the gravitational constant, $c$ is the speed of light, and $\fblr=1$ is the so-called virial factor that
includes all the unknown information about the geometry and kinematics of the BLR gas. 
The dimensionless accretion rate $\dot{\mathscr{\cal M}}$ 
is estimated via (e.g., \citealt{Du2014,Lu2016a})
\begin{equation}
\mathdotM=20.1\left(\frac{\ell_{44}}{\cos i}\right)^{3/2}M_7^{-2}, 
\label{eqn_mdot}
\end{equation} 
where, $\ell_{44}=L_{5100}/10^{44}\ergs$ is the 5100 \AA\, luminosity, 
$M_7=\bhm/10^7\sunm$ is the black hole mass, $\cos i$ is the cosine of the inclination of the accretion disk. 
Following the usual approximation, we take $\cos i=0.75$. 
The fluxes of broad hydrogen Balmer lines (H$\beta^{\rm b}$, H$\alpha^{\rm b}$), narrow hydrogen Balmer lines (H$\beta^{\rm n}$, H$\alpha^{\rm n}$), [O\,III] $\lambda5007$ and [N\,II] $\lambda6583$ are measured from the best fitted model. 

\begin{figure*}
\centering
\includegraphics[angle=0,width=1.0\textwidth]{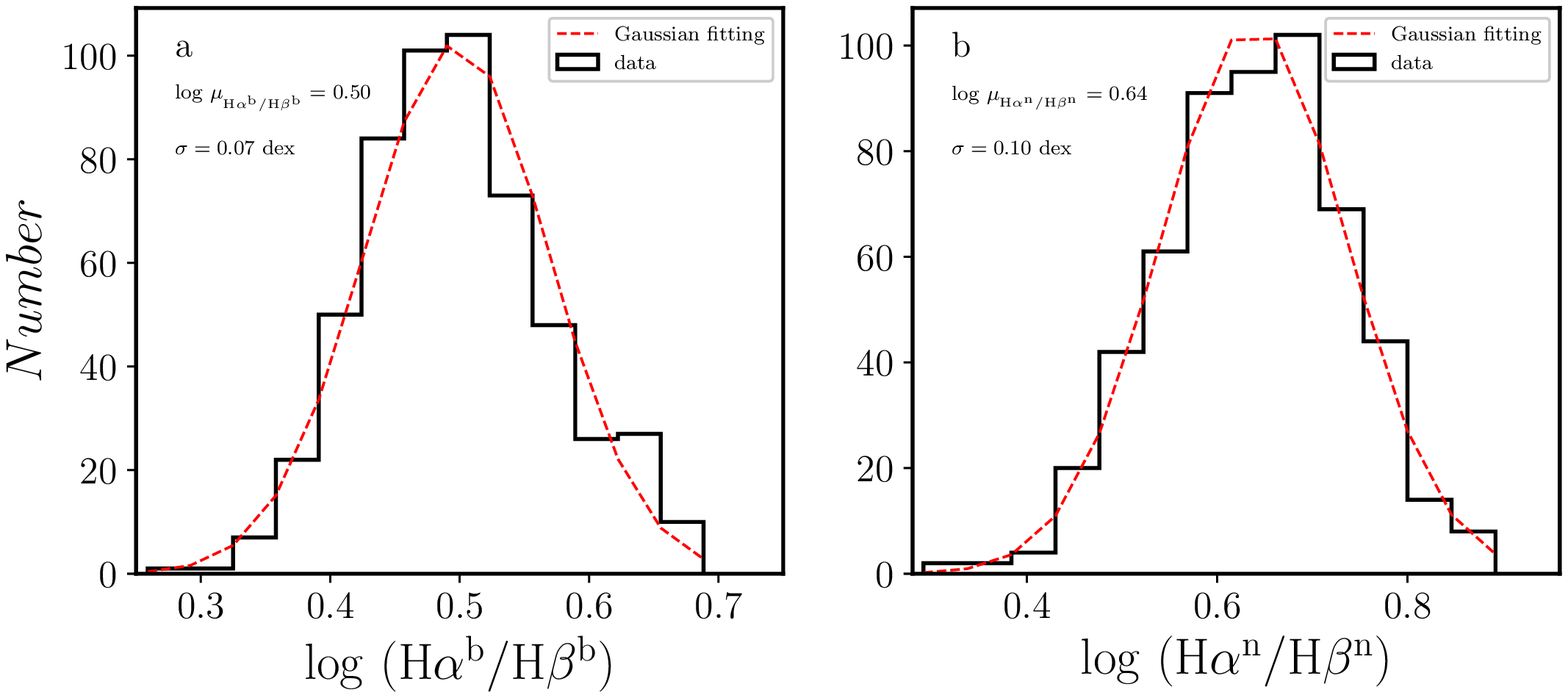}
\caption{
The distributions of broad- and narrow-line Balmer decrements, respectively. 
Panel ({\it a}) shows the histogram of $\bcd$ ratio in base-10 logarithm form, 
and panel ({\it b}) is the histogram of $\bnd$ ratio. 
In panel ({\it a, b}), we also noted the average value ($\mu$) and standard deviation of distributions. 
}
\label{fig-bd-gauss}
\end{figure*}

\begin{figure*}
\centering
\includegraphics[angle=0,width=1.0\textwidth]{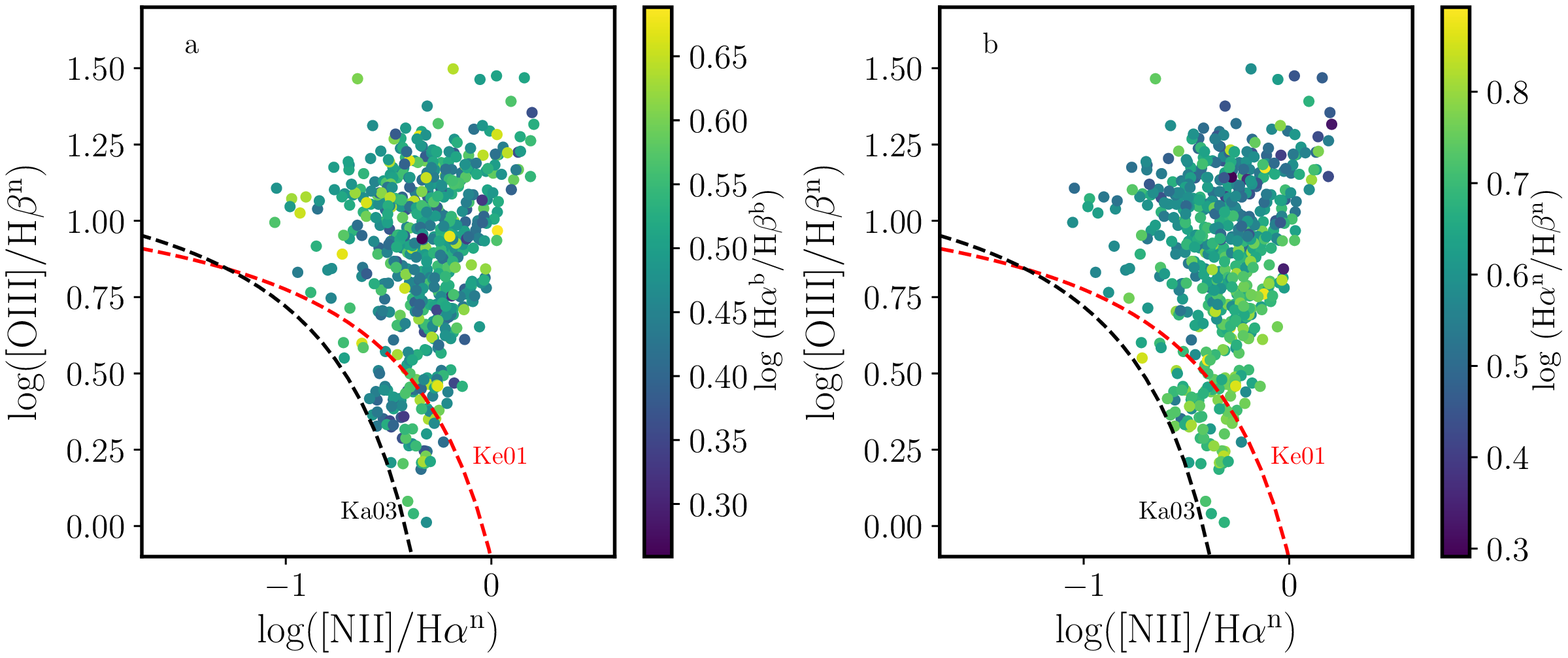}
\caption{
The 2D distributions of BPT$-$Balmer decrements. 
Panel ({\it a}) is BPT$-$ $\bcd$ diagram, 
panel ({\it b}) is BPT$-$ $\bnd$ diagram.  
The Ke01 line is the theoretical predictions of the scenarios from Kewley et al. (2001), 
and the Ka03 line is defined by Kauffmann et al. (2003). 
}
\label{fig-bd-bpt}
\end{figure*}

\section{Results}
\label{sec-r}
\subsection{The Distribution of Balmer Decrements}
\label{sec-1}
Using the Balmer line (H$\beta$ and H$\alpha$) fluxes, we calculated 
the broad-line Balmer decrements of $\bcd$ and the narrow-line Balmer decrements of $\bnd$ for the 554 AGNs. 
In Figure~\ref{fig-bd-gauss}({\it a, b}), we plotted the distributions of $\bcd$ and $\bnd$ ratios 
in logarithm form by dividing the samples into 14 bins, respectively. 
Both distributions can be fitted by a Gaussian function, as shown with the red dashed line in Figure~\ref{fig-bd-gauss}.  
The best-fitted results give the mean log($\bcd$)=0.50 (i.e. $\bcd$=3.16) 
with a standard deviation of 0.07 dex, and the mean log($\bnd$)=0.64 (i.e. $\bcd$=4.37), with a standard deviation of 0.10 dex. 
These average ratios are respectively larger than the intrinsic values of $\bcd=2.72$ (refer to \citealt{Gaskell2017} for this value) and $\bnd=3.1$ \citep{Gaskell1982,Gaskell1984,Wysota1988}.  
The average ratios of $\bcd$ and $\bnd$ strongly show that the narrow-line Balmer decrements are much larger than 
broad-line systematically. This suggests that the level of reddening in the NLR is higher than the BLR 
if we accept that the observed Balmer decrements depend on the reddening. 

\subsection{Balmer Decrements and BPT diagram}
\label{sec-bpt} 
The two-dimensional line-intensity ratio calculated from relatively strong lines of 
[O~III]~$\lambda$5007, H$\beta~\lambda$4861, [N~II]~$\lambda$6584 and H$\alpha~\lambda$6563 
(here Balmer emission-line refer only to the narrow component) 
can be used to probe the nebular conditions of a source (i.e., BPT diagram promoted by \citealt{Baldwin1981} and modified by \citealt{Veilleux1987}). 
Using Balmer decrements as third parameter, 
we constructed 2D distributions of BPT$-$Balmer decrement for 554 AGNs in Figure~\ref{fig-bd-bpt}({\it a, b}), 
and investigated the relationship between Balmer decrements and physical condition of nebular. 
We found that only 35 (6\%) of 554 AGNs lie below Ke01 line \citep{Kewley2001}, 
which means that the ionization of the NLR is dominated by AGN rather than host galaxy. 

In Figure~\ref{fig-bd-bpt}~({\it a}), we can see that $\bcd$ distribute in the BPT diagram randomly, 
which is confirmed by the correlation analysis (see Table~\ref{tab-1}). 
Panel ({\it b}) of Figure~\ref{fig-bd-bpt} shows that most of the AGNs with the largest $\bnd$ ratios 
locate on the bottom of the BPT diagram (the direction of [O~III]/H$\beta^{\rm n}$ axis), 
which in fact indicates a correlation between $\bnd$ and [O~III]/H$\beta^{\rm n}$. 
This is confirmed by the correlation analysis, with correlation coefficient $r_{\rm s}=-0.55$, 
and probability of no correlation $P_{\rm null}=1.80\times10^{-45}$. 
These results indicate that the broad-line Balmer decrements are independent on the physical conditions of the NLR, 
whereas the narrow-line Balmer decrements are associated with the physical conditions of the NLR. 
We will further discuss this in the following analysis.

\begin{figure}
\centering
\includegraphics[angle=0,width=0.5\textwidth]{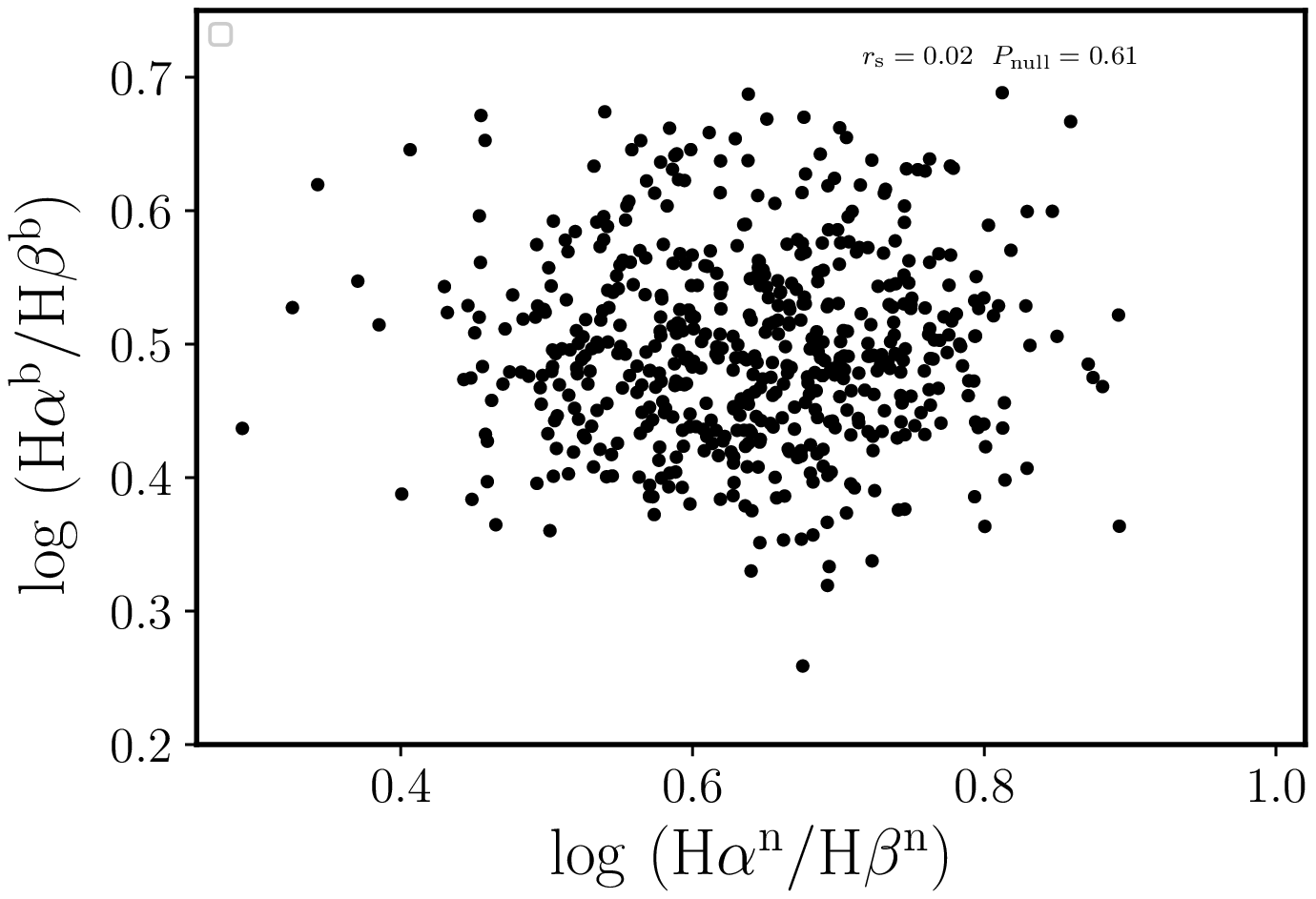}
\caption{
The relationship between $\bcd$ and $\bnd$.
$r_{\rm s}$ and $P_{\rm null}$ are the Spearman rank correlation coefficient 
and probability of the null hypothesis, respectively. 
}
\label{fig-bbd-nbd}
\end{figure}

\begin{figure*}
\centering
\includegraphics[angle=0,width=1.0\textwidth]{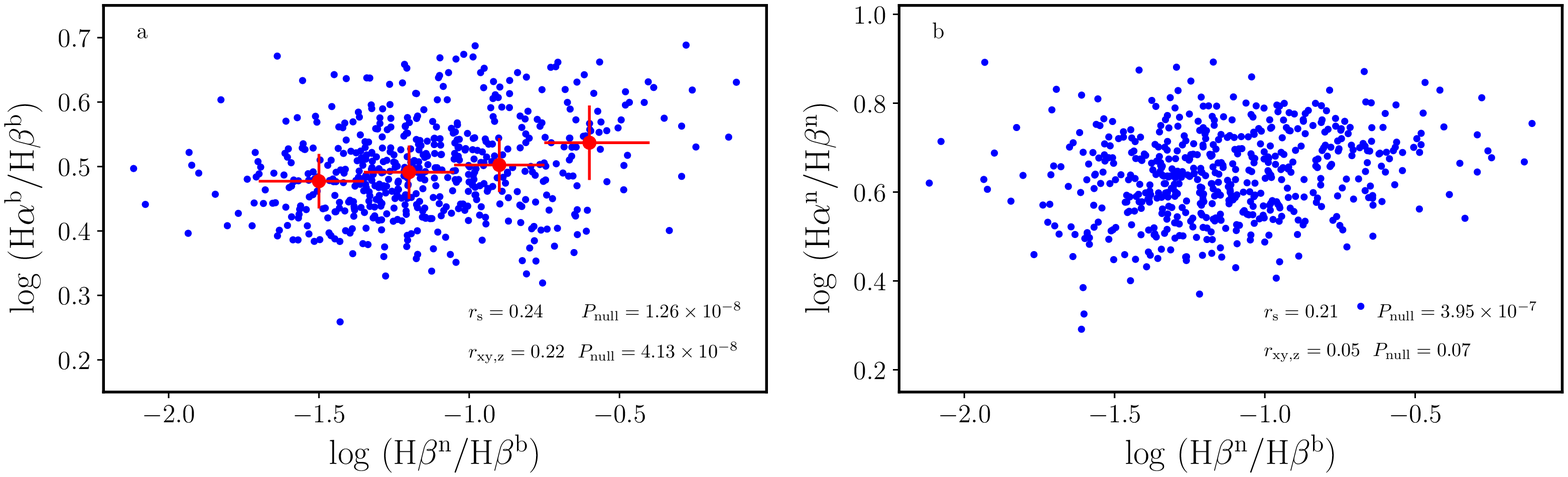}
\caption{
The relationship between Balmer decrement ($\bnd$ and $\bcd$) and H$\beta^{\rm n}$/H$\beta^{\rm b}$. 
The results of correlation analysis are quoted in panels. 
Splitting H$\beta^{\rm n}$/H$\beta^{\rm b}$ ratios in logarithm into 4 bins, 
and estimating the median value of $\bcd$ in each bin, we plotted the results (red dots) in panel~({\it a}). 
Whereas partial correlation analysis shows that there is no correlation between $\bnd$ and H$\beta^{\rm n}$/H$\beta^{\rm b}$, 
we did not bin them in panel~({\it b}). 
}
\label{fig-bd-inc}
\end{figure*}

\subsection{Relationship of $\bcd$ and $\bnd$}
\label{sec-cor}
In this section, we attempt to investigate the relationship between $\bcd$ and $\bnd$ for our sample, 
as shown in Figure~\ref{fig-bbd-nbd}. 
The Spearman rank correlation coefficient $r_{\rm s}$ 
and probability of the null hypothesis $P_{\rm null}$ (listed in Table~\ref{tab-1}) are $0.02$ and $0.61$ respectively, 
indicating that $\bnd$ does not correlate with $\bcd$, 
which is consistent with the find of \cite{deZotti1985}. 
The dust causing the common reddening of both the BLR and NLR could be associated with the dust of host galaxy \citep{deZotti1985,Baron2016}, because host galaxy is extremely extended. 
However, if reddening of the BLR and NLR is caused fully by the same dust in the line of sight, 
the broad-line Balmer decrements should be significantly correlated with the narrow-line Balmer decrements, vice versa. 
We found that the averaged $\bcd$ is smaller than the averaged $\bnd$ (Section~\ref{sec-1}), 
and they are independent on each other. 
These indicate that the NLR could be covered by more dust in the light of sight, 
and the dust causing the reddening of the BLR and NLR may be different. 
Many cases can account for this result. For example, 
in the frame of AGN unified model (\citealt{Antonucci1993}), 
it is possible that the partial dusty torus of AGN cause the reddening of the BLR, 
and the dust amount obscuring the BLR in the line of sight depend on the inclination angle of AGN. 
But the NLR may be free from the dust torus 
since the NLR gas lies under the gravitational influence of host galaxy (e.g., \citealt{Ho2009}). 
However, as suggested by \cite{Mor2009}, \cite{Stern2014}, \cite{Baron2016} and \cite{Ramos2017}, 
the NLR is dusty clouds, which may cause the reddening of the NLR in observation. 
Certainly, it is also possible that both the BLR and NLR may be reddened by the dust in the host galaxy, 
and thus the total dust amount sustained by the BLR and NLR are dependent on their sizes and the observing angle.

\begin{table}
\centering
\caption{
Summary of the Spearman rank correlation coefficient 
$r_{\rm s}$ and probability of the null hypothesis $P_{\rm null}$. 
}
\label{candidates}
\begin{tabular}{lccccccccrccccccc}
\hline
Parameter 1 & Parameter 2 &  $r_{\rm s}$  & $P_{\rm null}$ \\
\hline
$\bnd$ & $\bcd$                                                             & 0.02  & $0.61$ \\
$\bcd$ & $\rm [N~II]~\lambda$6584$\rm/H\alpha^{n}$   & 0.02 & $0.60$ \\
$\bcd$ & $\rm [O~III]~\lambda$5007$/\rm H\beta^{n}$  & 0.06 & $ 0.19$ \\
$\bnd$ & $\rm [N~II]~\lambda$6584$\rm/H\alpha^{n}$  & 0.02  & $ 0.68$ \\
$\bcd$ & $\rm H\beta^{n}/\rm H\beta^{b}$ &  0.24  & $1.26 \times 10^{-8}$ \\
$\bnd$ & $\rm H\beta^{n}/\rm H\beta^{b}$ &  0.21  & $3.95 \times 10^{-7}$ \\  
$\bcd$ & $L_{5100}$                         & -0.10 & 0.01 \\
$\bnd$ & $L_{5100}$                          & -0.01 & 0.75 \\
$\bcd$ & $\dot{\mathscr{\cal M}}$    & -0.18 & $2.47 \times 10^{-5}$ \\
$\bnd$ & $\dot{\mathscr{\cal M}}$    & 0.24 & $6.16 \times 10^{-9}$ \\
$\bnd$ & $\rm [O~III]~\lambda$5007$/\rm H\beta^{n}$      & -0.55  & $1.80 \times 10^{-45}$ \\
$\bnd$ & EW($\rm [O~III]$)      & -0.59  & $2.04 \times 10^{-52}$\\
$\rm [O~III]~\lambda$5007$/\rm H\beta^{n}$  & $\dot{\mathscr{\cal M}}$  & -0.33  & $3.41 \times 10^{-15}$ \\
EW($\rm [O~III]$)      & $\dot{\mathscr{\cal M}}$  & -0.25  & $2.46 \times 10^{-9}$\\
\hline
\end{tabular}
\label{tab-1}
\end{table}

\subsection{Balmer Decrements and Seyfert subtype}
\label{sec-st} 
In order to investigate the relation between Balmer decrements and Seyfert sub-type. 
We chose the most physical scheme used by \cite{Osterbrock1977}, 
which subdivides the AGN population into Seyfert 1.0, 1.2, and 1.5 according to the prominence of 
the broad compared to the narrow component of H$\beta$ (i.e, H$\beta^{\rm n}$/H$\beta^{\rm b}$ ratio). 
That is, in the frame of AGN unified model \citep{Antonucci1993}, 
the higher H$\beta^{\rm n}$/H$\beta^{\rm b}$ ratio means the AGN with larger inclination angle to the telescope. 
However, \cite{Stern2012b} found that the NLR to BLR luminosity ratio (${\rm H\alpha}^{\rm n}/{\rm H\alpha}^{\rm b}$) 
decreases with increasing H$\alpha^{\rm b}$ luminosity ($L_{\rm H\alpha^{b}}$), 
and suggested that the drop of ${\rm H\alpha}^{\rm n}/{\rm H\alpha}^{\rm b}$ with $L_{\rm H\alpha^{b}}$ is most likely due to 
the decrease in covering factor of the NLR with AGN luminosity (also see \citealt{Ludwig2009}).
Therefore, 
in the next, we test the relationship between Balmer decrements and H$\beta^{\rm n}$/H$\beta^{\rm b}$ 
by removing the influence of H$\beta^{\rm b}$ luminosity ($L_{\rm H\beta^{\rm b}}$) 
using the partial correlation analysis \citep{Kendall1979,Lu2016b}. 

Figure~\ref{fig-bd-inc}~({\it a, b}) shows the broad- and narrow-line Balmer decrement 
as a function of H$\beta^{\rm n}$/H$\beta^{\rm b}$ ratio. 
We employed correlation analysis to investigate the dependency of broad- and narrow-line 
Balmer decrement on H$\beta^{\rm n}$/H$\beta^{\rm b}$ ratio. 
The Spearman rank correlation coefficient $r_{\rm s}$ 
and probability of the null hypothesis $P_{\rm null}$ are quoted in panel ({\it a, b}) 
and also listed in Tabel~\ref{tab-1}. 
The result shows that the broad-line Balmer decrements weakly correlate with 
H$\beta^{\rm n}$/H$\beta^{\rm b}$ ratio ($r_{\rm s}=0.24$, $P_{\rm null}=1.26\times10^{-8}$), 
and the narrow-line Balmer decrements weakly correlate with 
H$\beta^{\rm n}$/H$\beta^{\rm b}$ ratio ($r_{\rm s}=0.21$, $P_{\rm null}=3.95\times10^{-7}$).  
Meanwhile, we found that H$\beta^{\rm n}$/H$\beta^{\rm b}$ ratio decreases with increasing 
$L_{\rm H\beta^{b}}$ in our sample 
($r_{\rm s}=-0.63$, $P_{\rm null}=8.68\times10^{-63}$), 
which is consistent with previous result 
(see Figure 3 of \citealt{Stern2012b}, who adopt ${\rm H\alpha}^{\rm n}/{\rm H\alpha}^{\rm b}$ 
ratio and $L_{\rm H\alpha^{\rm b}}$). 
Therefore, we further employed partial correlation analysis 
to investigate whether these correlations are affected by $L_{\rm H\beta^{\rm b}}$.
The partial correlation coefficients show that $\bcd$ still weakly correlates with H$\beta^{\rm n}$/H$\beta^{\rm b}$ ratio 
($r_{\rm xy, z}=0.22$, $P_{\rm null}=4.13\times10^{-8}$, 
where $x \equiv {\rm H}\beta^{\rm n}/{\rm H}\beta^{\rm b}$, $y \equiv \bcd$ and $z \equiv L_{\rm H\beta^{\rm b}}$), 
whereas there is no correlation between $\bnd$ and H$\beta^{\rm n}$/H$\beta^{\rm b}$ 
($r_{\rm xy, z}=0.05$, $P_{\rm null}=0.07$, where $y \equiv \bnd$). 
The former trend is supported by the fact that the $\bcd$ of Seyfert 1.8-1.9s is 
above 10 and might be as high as 20 (\citealt{Osterbrock1981,Crenshaw1988,Osterbrock2006}). 
The lack of correlation between $\bnd$ and H$\beta^{\rm n}$/H$\beta^{\rm b}$ indicates that the inclination has little effect on the reddening of NLR emission lines, which might suggest that the reddening of NLR seems to be mainly caused by the dust embedded in the NLR clouds. 
 
\begin{figure*}
\centering
\includegraphics[angle=0,width=1.0\textwidth]{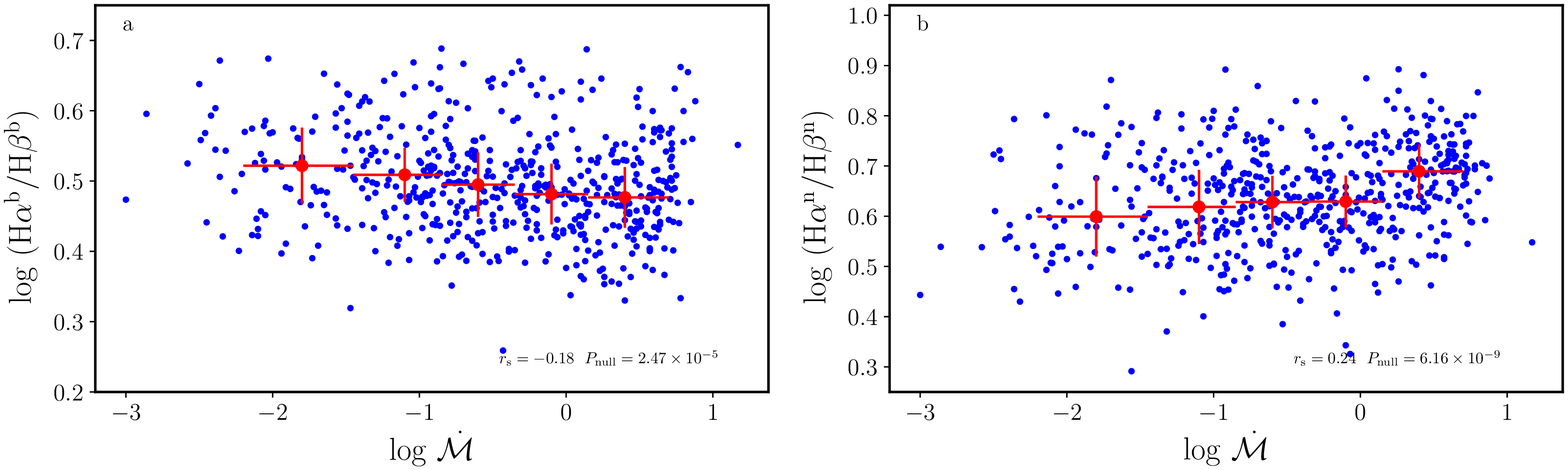}
\caption{
Same as Fig.~\ref{fig-bd-inc}, but for the relationship between Balmer decrement ($\bnd$ and $\bcd$) and accretion rate $\dot{\mathscr{\cal M}}$. 
}
\label{fig-bd_ar}
\end{figure*}

\subsection{Balmer Decrements and Optical Luminosity}
\label{sec-ol}
To check the dependency of Balmer decrements on optical luminosity, 
we employed the Spearman rank correlation analysis. 
The Spearman rank correlation coefficient of $r_{\rm s}$ and 
probability of the null hypothesis of $P_{\rm null}$ are also given in Table~\ref{tab-1}. 
The test shows that broad-line Balmer decrements do not correlate with optical luminosity, 
which is consistent with the result of \cite{Dong2008}. 
This suggests that dust covering factor in the BLR remains a constant with optical luminosity increasing. 
This case is opposite with receding torus models suggested that obscured fraction in the line of sight 
decreases with optical luminosity increasing \citep{Lawrence1991}.  
However, recent studies Based on sophisticated dusty torus model suggested that the dust covering factor weakly depends on AGN luminosity \citep{Stalevski2016}. 
Meanwhile, we found that the narrow-line Balmer decrement is also independent on optical luminosity of AGNs. 

Infrared (IR) time-lag observations show that the inner dust torus radius scales with the AGN luminosity as $r \propto L^{1/2}$ (``size-luminosity'' relation, \citealt{Suganuma2006}), 
as well as from theoretical calculation \citep{Barvainis1987} and near-IR(NIR) interferometric observations \citep{Kishimoto2011}. 
However, based on mid-IR(MIR) observational study of 23 AGNs, \cite{Burtscher2013} suggest that a common ``size-luminosity'' relation does not exist for AGN torus 
due to the fact that the torus sizes of these fainter sources are smaller than that expected from $r \propto L^{1/2}$ relation and show a large scatter, 
and two torus model components are needed to explain the observed results. 
The polar component was detected directly in NGC 1068 using high-angular-resoluiton MIR observations \citep{{Mason2006}}. 
Recently, the studies of MIR observation for $\sim$20 AGNs indicate that the bulk of the MIR emission comes from a diffuse polar component \citep{Asmus2016}. 
These results show that the nature of the nuclear environment of accreting systems is extremely complex. 
Our result suggests that the dust in the line of sight causing the reddening of the BLR and NLR 
do not be impacted by radiation pressure (i.e., optical luminosity) of accretion disk. 
This may caused by the fact that the intrinsic differences in their dust structures. 
It is also possible that the dust near the central region is destroyed (\citealt{MacAlpine1985,Osterbrock2006}),  
and the dust causing the reddening of AGN survives at the radius far from the central region of AGN, 
which attenuates the role of AGN optical luminosity on Balmer decrements.

\subsection{Balmer Decrements and accretion rate}
\label{sec-ar}
\cite{Ricci2017} found that  the covering factor of dusty torus drops dramatically with increasing $L/L_{\rm Edd}$. 
This could be a natural result that a part of dusty torus supply gas for accretion flow, 
and a part of dusty torus in high-latitude may drop to the low-latitude of accretion disk.
In this case, AGNs reddening (especially, for the region within the dusty torus) will be weakened with increasing accretion rate 
if the BLR (and accretion disk) are obscured by dusty torus. 
Therefore, broad-line Balmer decrement decreases as accretion rate increases. 
This expectation is supported by the test shown in Figure~\ref{fig-bd_ar}~({\it a}), 
where we plot Balmer decrements as a function of accretion rate, to some degree. 
The Spearman rank correlation analysis shows that broad-line Balmer decrements weakly correlate with accretion rate 
(see Table~\ref{tab-1} and Figure~\ref{fig-bd_ar}~{\it a}). 

Meanwhile, we found an interested results that the narrow-line Balmer decrement increases 
with increasing accretion rate (see Figure~\ref{fig-bd_ar}~{\it b}). 
Recall that the narrow-line Balmer decrements correlate with physical conditions of the NLR (Section~\ref{sec-bpt}), 
it is possible that the physical conditions in the NLR are driven/modulated by accretion rate. We will further investigate this issue in the next section.

\begin{figure*}
\centering
\includegraphics[angle=0,width=1.0\textwidth]{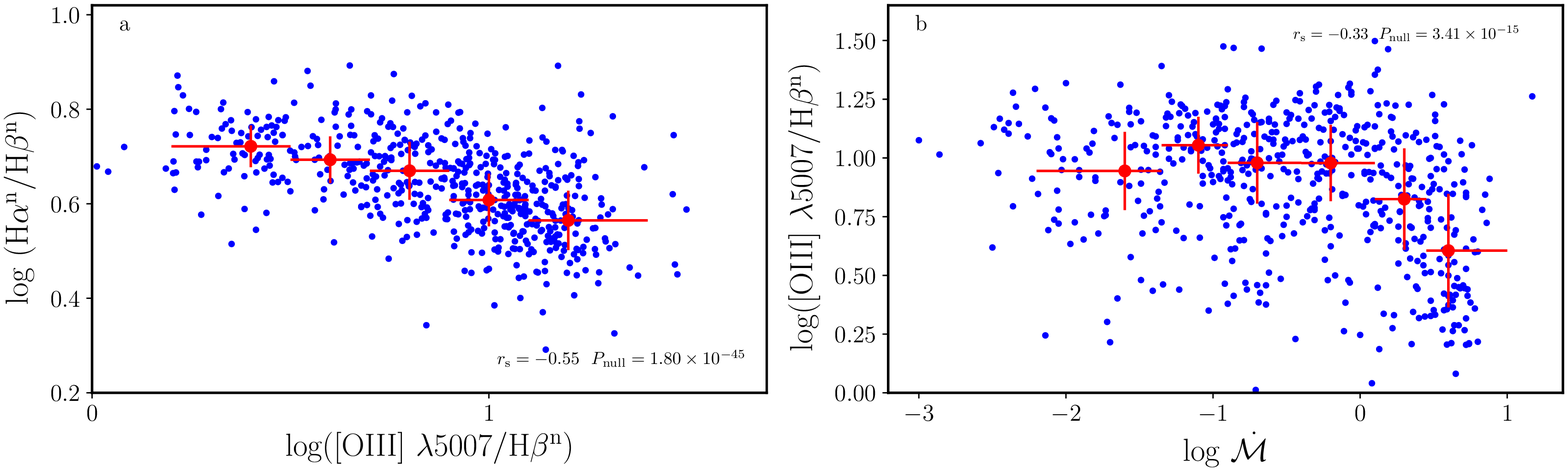}
\caption{
Same as Fig.~\ref{fig-bd-inc}, but for $\bnd-\rm [O~III]~\lambda5007/\rm H\beta^{n}-\dot{\mathscr{\cal M}}$ relationships. 
}
\label{fig-nbd-nlr-ar1}
\end{figure*}

\begin{figure*}
\centering
\includegraphics[angle=0,width=1.0\textwidth]{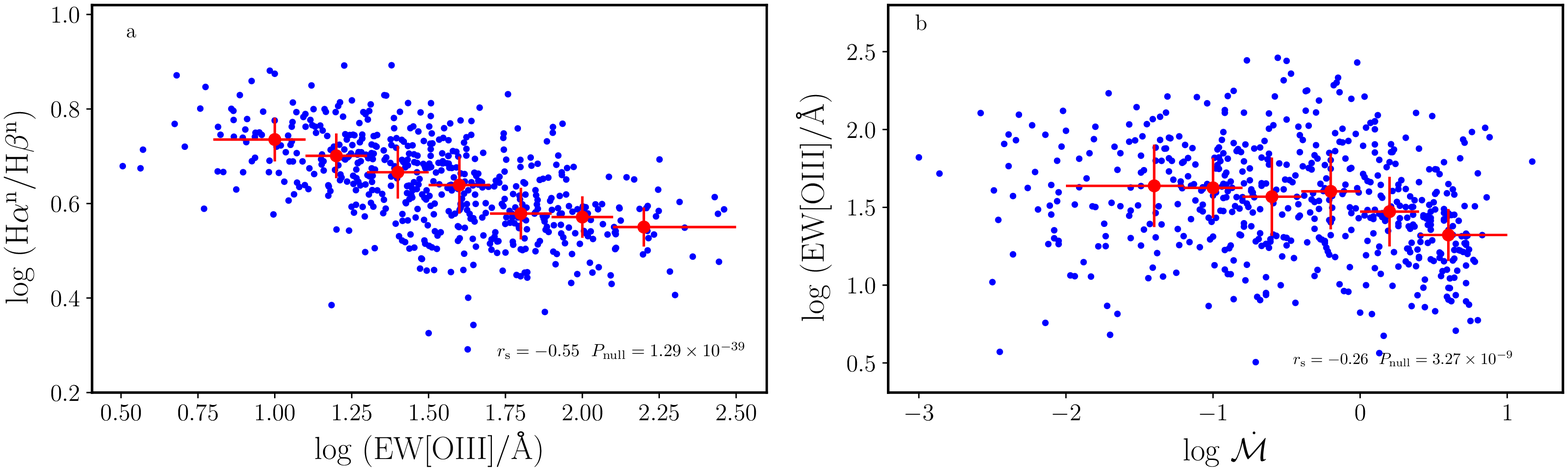}
\caption{
Same as Fig.~\ref{fig-bd-inc}, but for $\bnd-\rm EW([O~III])-\dot{\mathscr{\cal M}}$ relationships. 
}
\label{fig-nbd-nlr-ar2}
\end{figure*}

\subsection{Narrow-Line Balmer Decrements and NLR Physics}
We have found that narrow-line Balmer decrements are independent on the inclination of AGN (Section~\ref{sec-st}), 
but depend on the accretion rate (Section~\ref{sec-ar}) and physical conditions of the NLR (Section~\ref{sec-bpt}). 
Using photoionization code {\tt CLOUDY}, 
\cite{Baskin2005} calculated emission-line flux ratio of $\rm [O~III]~\lambda5007/\rm H\beta^{n}$, 
and found that $\rm [O~III]~\lambda5007/\rm H\beta^{n}$ depends on the electron density and 
ionization parameter of the NLR gas. 
Based on Palomar-Green quasar sample, \cite{Baskin2005} found that the strength of $\rm [O~III]~\lambda$5007 
is modulated by the covering factor,  electron density and ionization parameter of the NLR gas (Figure 5 of \citealt{Baskin2005}).  
Therefore, we use $\rm [O~III]~\lambda5007/\rm H\beta^{n}$ and EW($\rm [O~III]$) represent the physical conditions of the NLR 
to investigate the properties of narrow-line Balmer decrements in this section. 

Figure~\ref{fig-nbd-nlr-ar1} ({\it a}) displays $\rm [O~III]~\lambda5007/\rm H\beta^{n}-\bnd$ relationship. 
The result of correlation analysis (noted in panel and listed in Table~\ref{tab-1}) shows that 
$\bnd$ inversely correlates with $\rm [O~III]~\lambda5007/\rm H\beta^{n}$. 
Figure~\ref{fig-nbd-nlr-ar1} ({\it b}) displays $\dot{\mathscr{\cal M}}-\rm [O~III]~\lambda5007/\rm H\beta^{n}$ relationship, 
which shows that 
$\rm [O~III]~\lambda5007/\rm H\beta^{n}$ inversely correlates with accretion rate. 
This result indicates that narrow-line Balmer decrement increases with decreasing $\rm [O~III]~\lambda5007/\rm H\beta^{n}$, 
which could be attribute to the electron density increases and ionization parameter decreases 
of the NLR gas as accretion rate increases. 
In addition, we can see from Figure~\ref{fig-nbd-nlr-ar1} ({\it b}) that the 
observed $\rm [O~III]~\lambda5007/\rm H\beta^{n}$ ratio trends to have a maximum value $\sim$ 10. 
This observed ratio is consistent with the expectation of photoionization model, which give out the maximum 
$\rm [O~III]~\lambda5007/\rm H\beta^{n}$ ratio $\sim$ 10 (the top panel of Figure 2 in \citealt{Baskin2005}). 

Similarly, Figure~\ref{fig-nbd-nlr-ar2} ({\it a}, {\it b}) display $\rm EW([O~III])-\bnd$ and $\dot{\mathscr{\cal M}}-\rm EW([O~III])$ 
relationship, respectively. The results of correlation analysis show that $\rm EW([O~III])$ inversely correlates with $\bnd$, 
and $\dot{\mathscr{\cal M}}$ inversely correlates with $\rm EW([O~III])$. 
When the accretion rate increases, AGN outflows or winds might increase. However,  
there are no theoretical calculations or observations that support outflows or winds can reach to 
the NLR \citep{Netzer2006,Hickox2018}. 
That is the covering factor of the NLR might not increase with accretion rate. 
Therefore, the inverse correlation between $\rm EW([O~III])$ and accretion rate might indicates that the 
electron density increases and/or ionization parameter decreases as accretion rate increases.

\section{Discussion}
\label{sec-d}
\subsection{Comparison with Previous Work} 
Previous results from different AGN samples give different $\bcd$. 
For example, \cite{LaMura2007} derived $\bcd = 3.45\pm 0.65$ based on 90 Seyfert 1 galaxies from SDSS.  
\cite{Zhou2006} measured $\bcd$ ratios for 2000 narrow-line Seyfert 1 galaxies (NLS1s) 
and obtained the average $\bcd$ ratio is 3.028 with a dispersion of 0.36 (refer to Fig. 6 of \citealt{Zhou2006}). 
\cite{Dong2008} selected 446 low-redshift (z$\le$0.35) AGN 
in which the spectral slopes vary in the range from 1.5 to 2.7, 
the $\bcd$ ratios range from 2.3 to 4.2 and the distribution of 
the $\bcd$ ratios can be well described by log-Gaussian 
(Figure 3 of \citealt{Dong2008}), with a mean $\bcd$ ratio of 3.06. 
\cite{Gaskell2017} selected an extremely blue AGN sample with $\alpha_{\nu}>+0.2$ 
(where $f_{\nu} \varpropto \nu^{\alpha_{\nu}}$, $\alpha_{\nu}$ is measured from $\lambda$4030\AA~to $\lambda$5600\AA), 
these bluest AGNs have a significantly flatter Balmer decrement (see Fig. 1 of \citealt{Gaskell2017}), 
which give the geometric mean $\bcd$ ratio is 2.86$\pm$0.06. 
For our sample, the $\bcd$ ratios can be described 
by a Gaussian function with a peak value 3.16 and standard deviation 0.07 dex. 
\cite{Dong2008} shown that the Balmer decrement is a good indicator of internal reddening in AGN. 
\cite{Gaskell2017} gave a conclusion that the bluest 10\% of SDSS AGNs have significant reddening, 
and suggested that the un-reddening Balmer decrement of the BLR is the Case B value of $\bcd\approx2.72$. 
On the whole, the results obtained from early small samples and our sample 
point to the mean $\bcd$ ratio of the BLR is larger than the Baker-Menzel Case B value, 
suggesting that the observed BLR is reddened by the dust in the line of sight. 
In addition, the recombination theory gives the narrow-line Balmer decrement of $\bnd\approx 2.85$, 
the best overall average value of 3.1 is adopt for the intrinsic $\bnd$ ratio of the NLR 
by taking into account the collisional excitation in the partly ionized transition region (\citealt{Osterbrock2006}). 
For our sample, the $\bnd$ ratios of the NLR also can be described 
by a Gaussian function with a peak value of 4.37 and standard deviation 0.10 dex, 
which is significantly larger than the typical value of 3.1. 
Combining with the suggestion that the narrow-line Balmer decrements do give a reliable indication of reddening (\citealt{Gaskell1982}), 
we concluse that the NLR is also reddened by the dust in the line of sight. 

On the other hand, \cite{Heard2016} studied a compiled sample, 
which is composed of \cite{Dong2008} blue AGNs, 
\cite{Dong2005} partially obscured AGNs, very high S$/$N AGNs observed 
by \citealt{Osterbrock1977} and \citealt{Cohen1983} in Lick observatory, 
and found that the average broad-line Balmer decrement is larger than the average narrow-line Balmer decrement. 
However, this is opposite to the result obtained from 109 nearby Seyfert galaxies by \cite{deZotti1985}.  
For our selected AGN sample, 
the average $\bnd$ ratio is also larger than the $\bcd$ ratio, 
this is consistent with result of \cite{deZotti1985} and \cite{Baron2016}\footnote
{\cite{Baron2016} calculated the broad- and narrow-line Balmer decrements 
of 1296 AGNs in the redshift range $0.35 < z < 0.4$ (see Fig. 7 of \citealt{Baron2016}), 
it is obvious that the narrow-line Balmer decrement is larger than the broad-line Balmer decrement.}.  

\subsection{Spatial Distribution of Dust}
In Section~\ref{sec-r}, we investigated the physical properties of broad- and narrow-line Balmer decrements systematically,  
and discovered many differences between the broad- and narrow-line Balmer decrements. 
If we accept that the steeper Balmer decrement primarily depends on internal reddening of AGN, 
our results are an important clue to investigate the space distribution of dust causing internal reddening of AGN. 

A probable case is that the dust causing the reddening of observed AGN survive in the radius far from the centre of AGN 
resulting from the broad-line Balmer decrements are a constant with optical luminosity (equivalently radiation pressure). 
As suggested by \cite{Ramos2017}, 
these dust may include equatorial/toroidal structure (torus) and polar component. 
For the BLR, it is possible that partial dust torus may cause the reddening of the BLR, 
and covering fractions of dusty torus in the line of sight depend on the inclination angle and accretion rate of AGN, 
which is supported by the fact that the broad-line Balmer decrement correlates with 
H$\beta^{\rm n}$/H$\beta^{\rm b}$ ratio and accretion rate 
(see Figure~\ref{fig-bd-inc} {\it a} and Figure~\ref{fig-bd_ar} {\it a}) and by the unified model of AGN. 
But complicated structure of the dusty torus, such as clumpy geometry and broad range of covering factors, 
and the existence of pole dust \citep{Ramos2017}, where covering factors of torus marginally depend on the accretion rate, 
could be attenuate the correlation between broad-line Balmer decrements and H$\beta^{\rm n}$/H$\beta^{\rm b}$ ratio. 
For the NLR, there is no significant correlation between the broad- and narrow-line Balmer decrement, 
and averaged broad-line Balmer decrement is lower than the narrow-line, suggesting that 
there are more and different dust reddening the NLR compared to the BLR. 
Interestingly, we found that narrow-line Balmer decrements correlate with physical conditions of the NLR and accretion rate. 
It is possible that many dust indwell in the NLR and redden the NLR (e.g., \citealt{Mor2009,Stern2014,Baron2016,Ramos2017}). 
 
 It should be noted, \cite{deZotti1985} argued that the reddening in the BLR arises in the plane of the galaxies. 
 This argument is based on interpreting the Balmer decrement of the BLR slightly correlates with axial ratio, 
 but the significant level of this correlation is very low. 
 To interprete NaI D as interstellar absorption line in the host galaxies, and the correlation of
 equivalent width of NaI D with optical continuum slope ($\alpha_{\rm opt}$), 
 \cite{Baron2016} suggested that the dusty gas causing the AGN reddening is the interstellar medium (ISM) of the host galaxy. 
 As \cite{Gaskell2017} concerned, it is possible that the variation in equivalent widths of NaI D 
 is due to varying amounts of host-galaxy starlight. 

\section{Conclusion}
\label{sec-c}
In this paper, we selected an AGN sample and investigated many physical properties of broad- and narrow-line Balmer decrement. 
Our main results are: 
\begin{enumerate}
 \item 
 The distributions of the broad- and narrow-line Balmer decrements in our sample can be described by a Gaussian function, 
 and the average broad- and narrow-line Balmer decrements are 3.16 and 4.37, respectively. 
The narrow-line Balmer decrements are systematically larger than the broad-line decrements, 
which is contrary in part to previous results. Using Balmer decrements as indicator of internal reddening of AGN, 
we can conclude that the dust causing the reddening in the NLR is larger than the BLR. 
\item 
The broad- and narrow-line Balmer decrements are independent and dependent on 
the physical conditions of the NLR, respectively.  
The broad-line Balmer decrements do not correlate with narrow-line Balmer decrements 
(e.g., \citealt{deZotti1985,Heard2016}). 
\item 
Using H$\beta^{\rm n}$/H$\beta^{\rm b}$ ratio (Seyfert sub-type) as the proxy of inclination angle, 
We found that the broad-line Balmer decrements weakly correlate with the inclination of AGN, 
while narrow-line Balmer decrements are independent on the inclination. 
We also found that broad- and narrow-line Balmer decrements are 
independent on AGN optical luminosity in our samples, which is consistent with the results of \cite{Dong2008}. 

\item
The Balmer decrements of the BLR (NLR) inversely (positively) correlate with accretion rates, 
which probably indicates that the reddening of the BLR and NLR may be modulated by accretion rate. 
\end{enumerate}

\section{Acknowledgements}
We are grateful to the referee for constructive suggestions that significantly improved the manuscript. 
We acknowledge the support of the staff of the Lijiang 2.4m telescope. 
Funding for the telescope has been provided by CAS and the People's Government of Yunnan Province. 
YZ is supported by the National Key Research and Development Program of China (No. 2017YFA0402704). 
K.X.L.  is supported by the Light of West China Program provided by CAS (No. Y7XB016001), 
and the National Natural Science Foundation of China (grant Nos. 11703077).

\end{document}